\documentclass[conference]{IEEEtran}

\usepackage[utf8]{inputenc} 
\usepackage[T1]{fontenc}    
\usepackage{hyperref}       
\usepackage{url}            
\usepackage{booktabs}       
\usepackage{amsfonts}       
\usepackage{nicefrac}       
\usepackage{microtype}      
\usepackage{amssymb}        
\usepackage{bbm}   
\usepackage{amsmath}   
\usepackage{graphicx}  
\usepackage{epstopdf}  
\usepackage{algorithm}
\usepackage{algpseudocode}
\usepackage{wrapfig}

\begin{document}

\bibliographystyle{IEEEtran}

\title{Learn to Allocate Resources in Vehicular Networks}

\author{Liang~Wang,~\IEEEmembership{Member,~IEEE,}
        Hao~Ye,~\IEEEmembership{Student Member,~IEEE,}
        Le~Liang,~\IEEEmembership{Student Member,~IEEE,}
        and~Geoffrey~Ye~Li,~\IEEEmembership{Fellow,~IEEE}
\thanks{Liang Wang is with the Key Laboratory of Modern Teaching Technology, Ministry of Education, Xi'an 710062, China, and the School of Computer Science, Shaanxi Normal University, Xi'an 710119, China (e-mail: wangliang@snnu.edu.cn).}

\thanks{Hao Ye, Le Liang and Geoffrey Ye Li are with the Department
of Electrical and Computer Engineering, Georgia Institute of Technology, Atlanta,
GA, 30332 USA (e-mail: \{yehao, lliang\}@gatech.edu; liye@ece.gatech.edu).}
}

\author{\IEEEauthorblockN{Liang Wang}
\IEEEauthorblockA{\textit{School of Computer Science} \\
\textit{Shaanxi Normal University}\\
Xi’an, China \\
wangliang@snnu.edu.cn}
\and
\IEEEauthorblockN{Hao Ye}
\IEEEauthorblockA{\textit{Department
of Electrical and Computer Engineering} \\
\textit{Georgia Institute of Technology}\\
Atlanta, USA \\
yehao@gatech.edu}
\and
\IEEEauthorblockN{Le Liang}
\IEEEauthorblockA{\textit{Department
of Electrical and Computer Engineering} \\
\textit{Georgia Institute of Technology}\\
Atlanta, USA \\
lliang@gatech.edu}
\and
\IEEEauthorblockN{Geoffrey Ye Li}
\IEEEauthorblockA{\textit{Department
of Electrical and Computer Engineering} \\
\textit{Georgia Institute of Technology}\\
Atlanta, USA \\
liye@ece.gatech.edu}
}

\author{\IEEEauthorblockN{Liang Wang}
\IEEEauthorblockA{\textit{School of Computer Science} \\
\textit{Shaanxi Normal University}\\
Xi’an, China \\
wangliang@snnu.edu.cn}
\and
\IEEEauthorblockN{Hao Ye, Le Liang, Geoffrey Ye Li}
\IEEEauthorblockA{\textit{Department
of Electrical and Computer Engineering} \\
\textit{Georgia Institute of Technology}\\
Atlanta, USA \\
\{yehao, lliang\}@gatech.edu; liye@ece.gatech.edu}
}
\maketitle

\begin{abstract}
Resource allocation has a direct and profound impact on the performance of vehicle-to-everything (V2X) networks. Considering the dynamic nature of vehicular environments, it is appealing to devise a decentralized strategy to perform effective resource sharing. In this paper, we exploit deep learning to promote coordination among multiple vehicles and propose a hybrid architecture consisting of centralized decision making and distributed resource sharing to maximize the long-term sum rate of all vehicles. To reduce the network signaling overhead, each vehicle uses a deep neural network to compress its own observed information that is thereafter fed back to the centralized decision-making unit, which employs a deep Q-network to allocate resources and then sends the decision results to all vehicles. We further adopt a quantization layer for each vehicle that learns to quantize the continuous feedback. Extensive simulation results demonstrate that the proposed hybrid architecture can achieve near-optimal performance. Meanwhile, there exists an optimal number of continuous feedback and binary feedback, respectively. Besides, this architecture is robust to different feedback intervals, input noise, and feedback noise.
\end{abstract}

\section{Introduction and Background}
Connecting vehicles on the roads as a dynamic communication network, commonly known as a vehicle-to-everything (V2X) network, is gradually becoming a reality to make our daily experience on wheels safer and more convenient \cite{16V2XService}. V2X enabled coordination among vehicles, pedestrians, and other entities on the roads can alleviate traffic congestion, improve road safety, in addition to providing ubiquitous infotainment services \cite{17V2XStandard,17V2PHY,19V2Net}. Recently, the 3rd generation partnership project (3GPP) begins to support V2X services in long-term evolution (LTE) \cite{SimScenario} and further the fifth
generation mobile communication system (5G) networks \cite{V2X5G}. Cross-industry alliance has also been founded, such as the 5G automotive association (5GAA), to push development, testing, and deployment of V2X technologies.

Due to high mobility of vehicles and complicated time-varying communication environments, it is very challenging to guarantee the diverse quality-of-service (QoS) requirements in vehicular networks, such as extremely large capacity, high reliability, and low latency \cite{19LowDelayEC}. To address such issues, efficient resource allocation for spectrum sharing becomes necessary in the V2X scenario.
Existing works on spectrum sharing in vehicular networks can be mainly categorized into two classes: centralized schemes \cite{18graphRA,19segMACV2X} and distributed approaches \cite{11DistCHAss,16DistV2V}. For the centralized schemes, decisions are usually made centrally at a given node, such as the head in a cluster or the base station (BS) in a given coverage area. In these schemes, the decision making node needs to acquire accurate channel state information (CSI), interference information of all the vehicle-to-vehicle (V2V) links, and each V2V link's transmit power to make spectrum sharing decisions. However, reporting all such information from each V2V link to the decision making node poses a heavy burden on the feedback links, and even becomes infeasible in practice.

As for distributed schemes \cite{11DistCHAss} \cite{16DistV2V}, each V2V link makes its own decision with partial or little knowledge of the transmission of other V2V links.
This may leave some channels overly congested while others underutilized, leading to substantial performance degradation. Inspired by the power of artificial intelligence (AI), especially reinforcement learning (RL) \cite{16AlphaGo}, the research community in wireless communications is gradually shifting the design paradigm to machine learning \cite{19RLfog}. In \cite{19MARL}, a multi-agent RL based spectrum sharing scheme is proposed to promote the payload delivery rate of V2V links while improving the sum capacity of vehicle-to-infrastructure (V2I) links.  In \cite{19HYeBroadcast}, each V2V link is treated as an agent to ensure the latency constraint while minimizing the interference to V2I link transmission. A dynamic reinforcement learning scheduling algorithm has been proposed to solve the network traffic and computation offloading problems in vehicular networks \cite{19RLoffload}.

In order to fully exploit the advantages of both centralized and distributed schemes while alleviating the requirement on CSI for spectrum sharing in vehicular networks, we propose a reinforcement learning-based resource allocation scheme with learned feedback as shown in Fig. \ref{Fig1}. In particular, we devise a centralized decision making and distributed spectrum sharing architecture to maximize sum rate of all links in the long run. In this architecture, each V2V link first observes the state of its surrounding channels and adopts a deep neural network (DNN) to learn what to feed back to the decision making unit, such as the BS, instead of sending all observed information directly. To maximize the long-term sum rate of all links, the BS then adopts deep reinforcement learning technique to allocate the spectrum for all V2V links. To further reduce feedback overhead, we adopt a quantization layer in each vehicle's DNN and learn how to quantize the continuous feedback.
The contributions of this paper are shown as below:
\begin{itemize}
  \item We combine the DNN and RL techniques to devise a centralized decision making and distributed spectrum sharing architecture for multiple V2V links in vehicular networks to maximize the long-term sum rate of all V2V links.
To reduce the feedback overhead while achieving the efficient spectrum sharing, each V2V link adopts a DNN to learn feedback information.
  \item To further reduce the feedback overhead and facilitate implementation, we employ a quantized layer for each vehicle's DNN and let each vehicle learn how to quantize the feedback.
\end{itemize}

The rest of this paper is organized as follows. The system model is presented in Section II. The centralized decision making and distributed spectrum sharing architecture is devised in Section III. Then, the qantized feedback scheme is proposed in Section IV. Simulation results are presented in Section V. Finally, conclusions are drawn in Section VI.

\section{System Model}

We consider a vehicular communication network with ${K}$ pairs of device-to-device (D2D) users and $N$ cellular users equipments (CUEs) coexisting with a BS, where all devices are equipped with a single antenna. In the V2X scenario, D2D users and CUEs can be vehicles or pedestrians. Each pair of D2D users exchange safety related messages \footnote{Without confusion, D2D pairs and V2V links are interchangeable in the rest of paper.} while each CUE uses a V2I link to support bandwidth-intensive applications, such as social networking and video streaming. In order to ensure the QoS of the CUEs, we assume all V2I links are assigned orthogonal radio resources. Let $\mathcal{K} = \left\{1,2,...,K \right\}$ and $\mathcal{N} = \left\{1,2,...,N \right\}$ denote the set of all D2D pairs and the set of all CUEs, respectively. Without loss of generality, we assume that each CUE occupies one channel for its uplink transmission. To improve the spectrum utilization efficiency, all V2V links share the spectrum resource with V2I links. Therefore, $\mathcal{N}$ is also referred to as the channel set.

We model the channel gain, $h_k^n$, between the transmitter and its corresponding receiver in the $k$-th D2D pair on the $n$-th channel as $h_{k}^n$. Similarly, we denote the channel gain from the $n$-th CUE to the BS on the $n$-th channel, i.e., the $n$-th V2I link, by $g_{n}$. Denote the cross channel from $n$-th CUE to the receiver of the $k$-th D2D pair on the $n$-th channel as $g^n_{k}$, and the cross channel from the transmitter of the $l$-th D2D pair to the receiver of the $k$-th D2D pair on the $n$-th channel as $h_{l,k}^n$.  Then, the data rate for the $k$-th D2D user on the $n$-th channel can be written as
\begin{equation}
r_k^n = B \log_2\left(1 + \frac{{\rho_k^n}  P_k \left| h_{k}^n \right|^2 }{\sum_{l \neq k}^{K}{{\rho_l^n}P_l \left| h_{l,k}^n \right|^2 }  + P_n \left| g^n_{k} \right|^2 + \sigma^2}\right),
\label{Eq2}
\end{equation}
where $B$ and $\sigma^2$ denote the channel bandwidth and the noise power respectively, $P_k$ and $P_n$ refer to the transmit powers of the $k$-th D2D pair and  the $n$-th V2I link, respectively. Besides,
 ${\rho_k^n} \in \left\{0,1 \right\}$ is the channel allocation indicator with ${\rho_k^n} = 1 $ if
the $k$-th D2D user pair chooses the $n$-th channel and ${\rho_k^n} = 0 $ otherwise. In addition, the terms $\sum_{l \neq k}^{K}{{\rho_l^n}P_l \left| h_{l,k}^n \right|^2 }$ and $P_n \left| g^n_{k} \right|^2$ in (\ref{Eq2}) refer to the interference of the remaining V2V links and the V2I link on the $n$-th channel, respectively.

In the V2X networks, a naive distributed approach will allow each D2D pair to perform channel selection such that its own data rate is maximized. However, local rate maximization often leads to suboptimal global performance. On the other hand, the BS in the V2X scenario has enough computational and storage resources to achieve efficient resource allocation. With the help of machine learning, we propose centralized decision making based on compressed information learned by each individual D2D pairs distributively.

In order to achieve this goal, each D2D pair first feeds back its learned state information related to the channel gain, the observed interference from other V2V links and V2I link, transmit power, etc. to the BS. Then, according to feedback information from all D2D pairs, the BS will make optimal decisions for all D2D users through reinforcement learning. Then, the BS will send the decision result to each D2D pair.

To limit overhead on information feedback, each D2D pair should only report the compressed information vector, $\mathbf{b}_k$, rather than all relevant information to the BS.  Here, $\left\{ b_{k,j} \right\}$ refers to the feedback vector of the $k$-th D2D and $b_{k,j}, \forall j \in \left\{1,2,...,{N}_k\right\}$ is the $j$-th feedback of the $k$-th D2D, where ${N}_k$ denotes the number of feedback learned by the $k$-th D2D pair . All D2D pairs aim at maximizing their global sum rate in the next transmission while minimizing the number of the feedback information $\mathbf{b}_k$.

\section{BS aided Spectrum Sharing Architecture}

We adopt the deep reinforcement learning approach for resource allocation in this section. We first discuss feedback information compression using a DNN at each D2D pair and then optimize decision making at the BS based on reinforcement learning, as shown in Fig. \ref{Fig1}.

\subsection{D2D Neural Network Design}
\begin{figure}
\begin{center}
\includegraphics[width=0.45\textwidth]{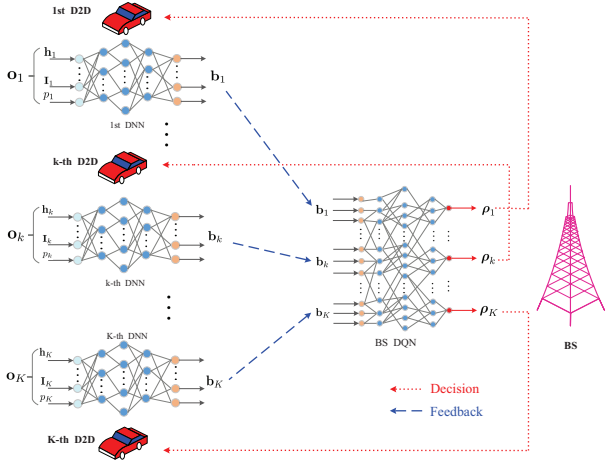}
\caption{Neural network architecture for the D2Ds and BS}
\label{Fig1}
\end{center}
\end{figure}

To fully explore the potentials of V2X networks and make best use of the computational and storage resources at the BS, we devise a new deep reinforcement learning based distributed information compression scheme for each V2V link while making decisions at the BS as in Fig. \ref{Fig1}. To determine the proper feedback, the transmitter of each V2V link utilizes the fully connected DNN to learn what to feed back. Particularly, the $k$-th D2D pair will first observe its corresponding surroundings and obtain the current channel state information and other related information, which is termed as local observation $\mathbf{o}_k$. As in Fig. \ref{Fig1}, $\mathbf{o}_k = \left\{\mathbf{h}_k, \mathbf{I}_k, p_k \right\}$, where $\mathbf{h}_k = \left(h_k^1,...,h_k^n,...,h_k^N \right)$ and $\mathbf{I}_k = \left(I_k^1,...,I_k^n,...,I_k^N \right)$. Here, $I_k^n$ refers the interference to the $k$-th D2D pair on the $n$-th channel, which can be expressed as $I_k^n = \sum_{l \neq k}^{K}{{\rho_l^n}P_l \left| h_{l,k}^n \right|^2 } + P_n \left| g^n_{k} \right|^2$. After that, the D2D user will treat the local observation $\mathbf{o}_k$ as input for the DNN and learn the feedback information $\mathbf{b}_k$ while maximizing their long-term sum rate at the BS globally. Here, note that the number of elements in $\mathbf{b}_k$ can be a variable and we should figure out its optimal value.

\subsection{BS Deep Q Network Design}
After designing the architecture of V2V link for distributed spectrum sharing, the deep Q network (DQN) architecture for the BS makes centralized decisions. In order to maximize the long-term sum rate of all links, we resort to the RL technique. We treat the BS as an agent. In order to allocate the proper spectrum to each D2D user pair, the BS treats all the feedback as the current state, $\mathcal{S}$, of the agent's environment, that is, $\mathcal{S} = \left\{\mathbf{b}_1,\mathbf{b}_2,...,\mathbf{b}_K \right\}$. Then, the actions of the BS is to determine the value of the channel indicators $\rho_k^n$. In other words, the action of the BS can be written as $\mathcal{A} = \left\{\boldsymbol{\rho}_1, ... , \boldsymbol{\rho}_k,  ..., \boldsymbol{\rho}_K \right\} ,\forall k \in \mathcal{K}$, where $\boldsymbol{\rho}_k = \left\{\rho_k^n \right\}, \forall n \in \mathcal{N}$ refers to the channel allocation vector for the $k$-th D2D pair. Finally, we model the reward of the BS as $R = \sum_{k=1}^{K}r_k = \sum_{k=1}^{K} \sum_{n=1}^N r_k^n $, where $r_k$ refers to the data rate of the $k$-th D2D pair on all the channels.

\subsection{Centralized Control and Distributed Transmission Architecture}

The overall centralized control and distributed transmission architecture is shown in Fig. \ref{Fig1}. Each V2V link first observes its transmission environment, such as, channel gain, interference and so on, and then adopts a DNN to compress this observed information into several real variables and finally feed this compressed information back to the BS. Using the feedback information of all V2V links as the input, the BS then utilizes DQN to perform  Q-Learning to decide which channel to choose for each V2V link, and finally send the channel selection decision to all V2V links.

Details of the training framework in Fig. \ref{Fig1} are provided in Algorithm \ref{Algo1}. In Algorithm \ref{Algo1}, we denote the estimation of the return also known as the approximate target value \cite{DQN} as
\begin{equation}
y_j = \sum_{k=1}^{K}r_k + \gamma \max\limits_{a'_j}{\mathcal{Q}\left(o'_j, a'_j, \boldsymbol{\theta}^{-}\right)},
\end{equation}
where $r_k$, $\gamma$, and ${\mathcal{Q}\left(o'_j, a'_j, \boldsymbol{\theta}^{-}\right)}$ are the reward of the $k$-th D2D pair, the discount factor, and the $\mathcal{Q}$ function of the target DQN with parameters $\boldsymbol{\theta}^{-}$ under the current observation $o'_j$ and action $a'_j$. Then, the updating process for the DQN of the BS can be written as \cite{DQN,16DoubleQN}:
\begin{equation}
\boldsymbol{\theta} \leftarrow \boldsymbol{\theta} + \beta \sum_{j \in \mathcal{D}} \frac{\partial {\mathcal{Q}\left(o_j, a_j, \boldsymbol{\theta} \right)}}{ \partial {\boldsymbol{\theta}}}\left[y_j - {\mathcal{Q}\left(o_j, a_j, \boldsymbol{\theta} \right)}\right] ,
\label{Eq13}
\end{equation}
where $\beta$ is the step size in one gradient iteration. Note that here we use $\mathbf{o} = \left\{ \mathbf{o}_k \right\}$ as the input of the whole neural network consisting of all D2Ds' DNNs and the BS DQN to implement an end-to-end training process. In addition, $N_u$ refers to the frequency that we copy the parameters $\boldsymbol{\theta}$ of BS DQN to the target DQN with parameters $\boldsymbol{\theta}^-$.

\begin{algorithm}
\caption{Training algorithm for the proposed architecture}
\hspace*{0.02in}{\bf Input:}
the DNN model for each D2D, DQN model for BS,
\hspace*{0.4in}V2X environment simulator \\
\hspace*{0.02in}{\bf Output:}
the DNN for each D2D, optimal control policy $\pi^*$
\hspace*{0.4in} represented by a DQN with parameters $\boldsymbol{\theta}$
\begin{algorithmic}[1]
    \State Initialize all DNNs and DQN models respectively
\For {episode $l = 1, ..., L$}
    \State Start the V2X environment simulator, generate vehicles,
    \item[] \hspace{\algorithmicindent}V2V links and V2I links
    \State Initialize the beginning observation $\mathbf{o}$ and the policy $\pi$ \item[] \hspace{\algorithmicindent}randomly
    \For {time-step $t = 1,...,T$}
        \State Each D2D adopts the observation $\mathbf{o}_t$ as the input
        \item[] \hspace{\algorithmicindent}\hspace{\algorithmicindent} of its DNN to learn the feedback $\mathbf{b}_k^t$
        \State BS takes $s_t = \left\{ \mathbf{b}_k^t \right\}$ as the input of its DQN,
        \item[] \hspace{\algorithmicindent}\hspace{\algorithmicindent} chooses $a_t$ from $\mathcal{A}$ using policy derived from $\mathcal{Q}$,
        \item[] \hspace{\algorithmicindent}\hspace{\algorithmicindent} i.e. $\epsilon$-greedy policy, and then broadcasts the action
        \item[] \hspace{\algorithmicindent}\hspace{\algorithmicindent} $a_t$ to every D2D
        \State Each D2D takes action based on $a$, get its reward
        \item[] \hspace{\algorithmicindent} \hspace{\algorithmicindent} $r_k^t$, the next observation $\mathbf{o}_{t'}$, and learn the next
         \item[] \hspace{\algorithmicindent} \hspace{\algorithmicindent} feedback $\mathbf{b}_k^{t'}$
        \State Save the data $\left\{\mathbf{o}_t, a_t, R_t,  \mathbf{o}_{t'} \right\}$ into the buffer $\mathcal{B}$
        \State Sample a mini-batch of data $\mathcal{D}$ from $\mathcal{B}$ uniformly
        \State Use the data in $\mathcal{D}$  to train the all D2Ds' DNNs and
        \item[] \hspace{\algorithmicindent} \hspace{\algorithmicindent} BS's DQN together as in Eq. (\ref{Eq13}).
        \State Each D2D updates the observation $\mathbf{o}_t \leftarrow \mathbf{o}_{t'}$, and
        \item[] \hspace{\algorithmicindent} \hspace{\algorithmicindent} the feedback  $\mathbf{b}_k^t \leftarrow \mathbf{b}_k^{t'}$
        \State Update target network: $\boldsymbol{\theta}^{-}\leftarrow \boldsymbol{\theta}$ every {$N_u$} steps
    \EndFor
\EndFor
\end{algorithmic}
\label{Algo1}
\end{algorithm}

\section{Spectrum Sharing with the binary feedback}

In order to further reduce the overheads of the feedback and facilitate the transmission in practical communication systems, we develop a framework to quantize the V2V links' real feedback into several binary data. In other words, we try to constrain ${b_{k,j}} \in \left\{-1,1 \right\},\forall k \in \mathcal{K}, \forall j \in \left\{1,2,...,{N}_k\right\}$.

The binary quantization process consists of two steps. The first step is to transform the learned continuous feedback into the continuous interval $\left[-1,1 \right]$. Then, taking the outputs of the first step as its input, the second step is to produce the desired number of the discrete outputs in the set $\left\{-1,1 \right\}$.

To implement the first step, we adopt a fully-connected layer with $tanh$ activations, where we term this layer as the pre-binary layer. Here, we have ${\rm tanh}\left( x\right) = \frac{2}{1 + e^{-2x}} - 1$. In order to quantize the continuous output of the first step, we adopt the traditional sign function method in the second step. To be specific, we take the sign of the input value as the output of this layer, which can be expressed as $b\left(x\right)$. However, the gradient of this function is not continuous, which is quite challenging considering the back propagation procedure while training the neural network in TensorFlow. As a remedy to this, we adopt the identity function in the backward pass, which is known as the straight-through estimator \cite{bengio2013estimating}.

Combining two steps together, the whole quantization process can be expressed as
$B\left( x \right) = b\left( {\rm tanh}\left(W_{0}x + b_{0}\right)\right)$, where $W_0$ and $b_0$ denote the linear weights and bias of the pre-binary layer that transform the activations from the previous layer in the neural network respectively.

\section{Simulation Results}
In this section, we conduct extensive simulation to verify the performance of the proposed scheme. Our simulation scenario is the urban case in Annex A of \cite{SimScenario}. The size of the simulation area is $1299$ m  $\times \; 750$ m, where the BS is located in the center of this area. We assume $N=K=4$, the carrier frequency, $f_c = 2$ GHz. The antenna height, antenna gain, and received noise figure of the BS are set as $25$ m, $8$ dBi and $5$ dB, respectively, while those of the vehicles are chosen as $1.5$ m, $3$ dBi and $9$ dB respectively. In addition, the vehicle drop and mobility model follows the urban case of A.1.2 in \cite{SimScenario}. The vehicle speed is randomly distributed within $[10, 15]$ km/h. Besides, the transmit powers of the V2I and the V2V links are $23$ dBm and $10$ dBm, respectively. The white noise is set as $\sigma^2 =  -114$ dBm. The channel model for V2I links $128.1 + 37.6 {\rm log}_{10}\left(d\right)$, where $d$ in km is the distance between the vehicle and the BS, while that of V2V link follows the LoS case in WINNER + B1 Manhattan in \cite{WinnerII}. The decorrelation distances of both link are $50$ m and $10$ m, respectively. The shadowing of the V2I and the V2V links follow Log-normal distribution with $8$ dB and $3$ dB standard deviation, respectively. The small-scale fading of both links follows Rayleigh distribution.
\begin{table}  
\small
\centering
\caption{Architecture for DNN and BS DQN}
\begin{tabular}{|p{2cm}|p{2.5cm}|p{2.5cm}|}
\hline
     & DNN & BS DQN \\
\hline
Input layer  &  9 & $K \times N_k$ \\
\hline
Hidden layers & 3 FC layers (16, 32, 16)  &   3 FC layers (1200, 800, 600)  \\  
\hline
Output layer  & $N_k$ &  $256$ \\
\hline
\end{tabular}
\label{TB1}
\end{table}
The architecture of the DNN for each D2D and BS DQN is shown in Table \ref{TB1}, where $N_{k}$ refers to the number of feedback and $FC$ denotes the fully connected (FC) layer, respectively. In addition, the number of neurons in its output layer is $256$, which refers to all the possible channel allocation for all V2V links. The rectified linear unit (ReLU) defined as $f\left(x\right) = \max{\left(0, x\right)}$, is chosen as the default activation function of the DNNs and the BS DQN in the proposed scheme. Here, the activation function of the output layers in the DNN and the BS DQN is set as the linear function. Besides, the RMSProp optimizer \cite{DLOptimizer} is adopted to update the network parameters with a learning rate of $0.001$. The loss function is set as Huber loss \cite{HuberLoss}. We use Keras \cite{chollet2015keras} for training and testing the proposed scheme, where TensorFlow is employed as the backend. The number of steps, $T=1,000$. The number of episodes in the training and testing periods is $2,000$. The update frequency $N_u$ of the target Q network is every $500$ steps. The discount factor $\gamma$ in the training is chosen as $0.05$. The size of the replay buffer $\mathcal{B}$ is set as $1,000,000$ samples. The mini-batch size $\mathcal{D}$ varies in different settings, which will be specified in each figure.

\begin{figure}[htbp]
\centering
\begin{minipage}[t]{0.45\textwidth}
\centering
\includegraphics[width=\textwidth]{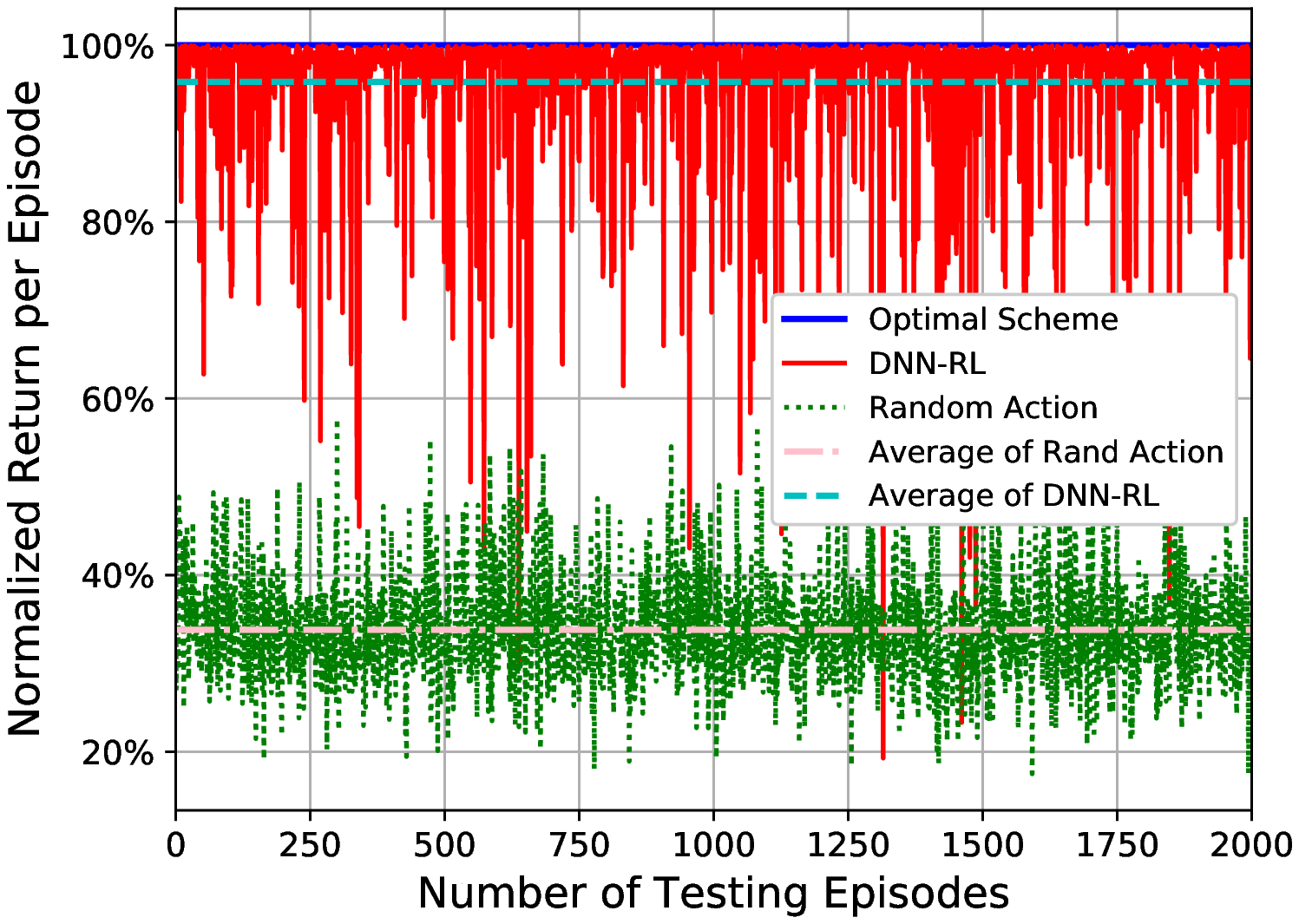}
\caption{Return comparison}
\label{Fig42}
\end{minipage}
\begin{minipage}[t]{0.45\textwidth}
\centering
\includegraphics[width=\textwidth]{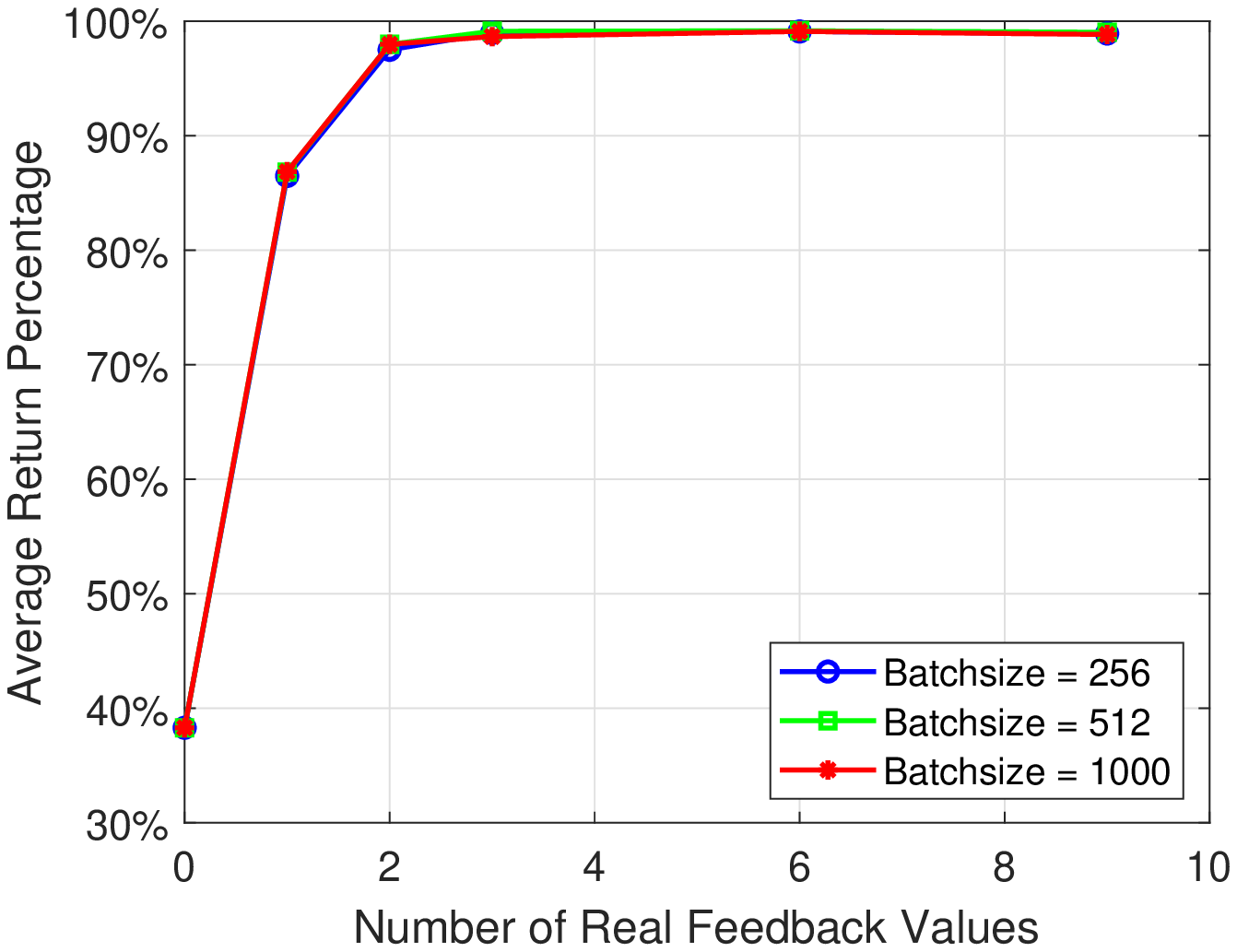}
\caption{ARP vs Real feedback}
\label{Fig5}
\end{minipage}
\begin{minipage}[t]{0.45\textwidth}
\centering
\includegraphics[width=\textwidth]{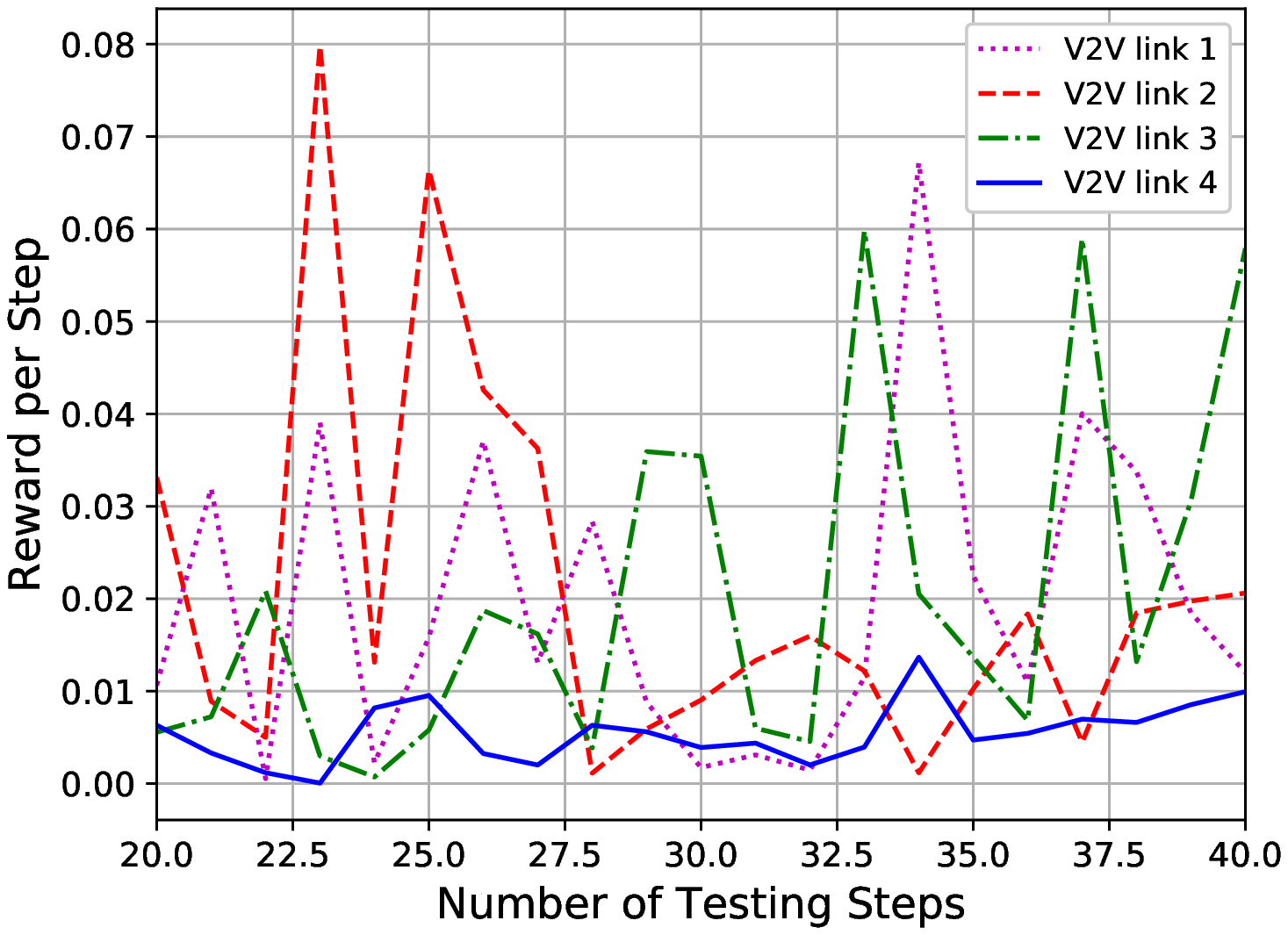}
\caption{Reward per V2V link}
\label{Fig43}
\end{minipage}
\end{figure}

Fig. \ref{Fig42} shows the return versus the number of testing episodes of the proposed DNN-RL scheme in all V2V links. Here, the mini-batch size $\mathcal{D}=512$ and the number of real feedback is $N_k=3$. For comparison, we also display the performance of other schemes. In the optimal scheme, we use brute-force search to find the optimal spectrum allocation in each testing step, which is very time consuming. In the random action scheme, each V2V link chooses the channel randomly. For better understanding, we depict the normalized return of these three schemes in Fig. \ref{Fig42}. Here, we use the return of the optimal scheme to normalize the other two schemes in each testing episode. Besides, the average values of our proposed scheme and the random action scheme are also depicted. In Fig. \ref{Fig42}, the performance of the optimal scheme is always $1$, while the performance of the DNN-RL approaches $1$ in many episodes and the average performance of the DNN-RL is about $95\%$ of the optimal scheme. But the average performance of random action is less than $40\%$ of the optimal performance. Thus, the proposed DNN-RL can achieve the near-optimal performance.

Fig. \ref{Fig5} shows the impacts of different batch sizes $\mathcal{D}$ and different numbers of real feedback on the performance of the DNN-RL scheme, which adopts the average return percentage (ARP) as the metric. Here, the ARP is defined as: the return under the DNN-RL is first averaged over $2,000$ episodes and then normalized by the average return of the optimal scheme.  In Fig. \ref{Fig5}, that the number of real feedback equals $0$ refers to the situation where V2V links do not feed anything back to the BS and therefore, the BS just randomly selects channel for each V2V link. From the figure, the ARP under the DNN-RL increases rapidly with the increasing number of real feedback, reaching the maximal percentage nearly $99\%$ at $3$ real feedbacks. Then, the ARP keeps nearly constant with the further increasing number of real feedbacks. In other words, each V2V link only needs to send $3$ real feedback values to the BS to achieve the near-optimal performance. From Fig. \ref{Fig5}, mini-batch size $\mathcal{D} = 512$ is good enough considering the computational overhead and the achieved performance.

Fig. \ref{Fig43} demonstrates the rate change, also known as reward per steps in RL terminology, of all V2V links at the $1200$-th testing episode. The rates of these four links change with testing steps due to the time-varying channels in the V2X scenario. For example, V2V link $3$ with the dash-dot line tends to have less data rate compared with V2V link $2$ with the dashed line in the first $23-27$ steps, but achieves a larger rate than V2V link $1$ in testing step $29-30$ and $33-34$ steps, which demonstrates that the proposed DNN-RL scheme can adapt to the time-varying channels in the V2X scenario. In addition, the rates of all V2V links are almost bigger than $0$, which shows that our proposed scheme can exploit the spectrum reuse property in the V2X scenario.

\begin{figure}[htbp]
\centering
\begin{minipage}[t]{0.45\textwidth}
\centering
\includegraphics[width=\textwidth]{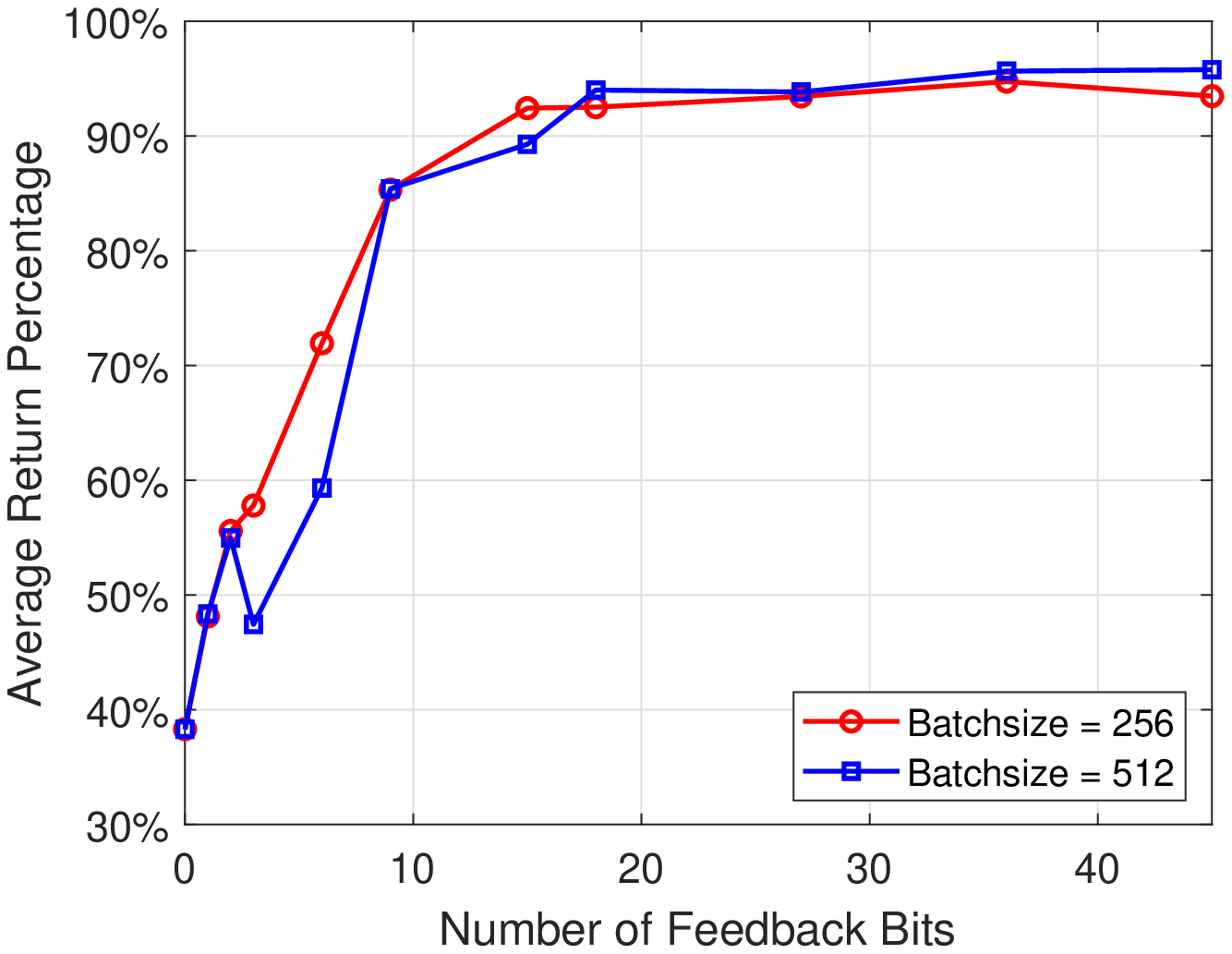}
\caption{ARP vs feedback bits}
\label{Fig6}
\end{minipage}
\begin{minipage}[t]{0.45\textwidth}
\centering
\includegraphics[width=\textwidth]{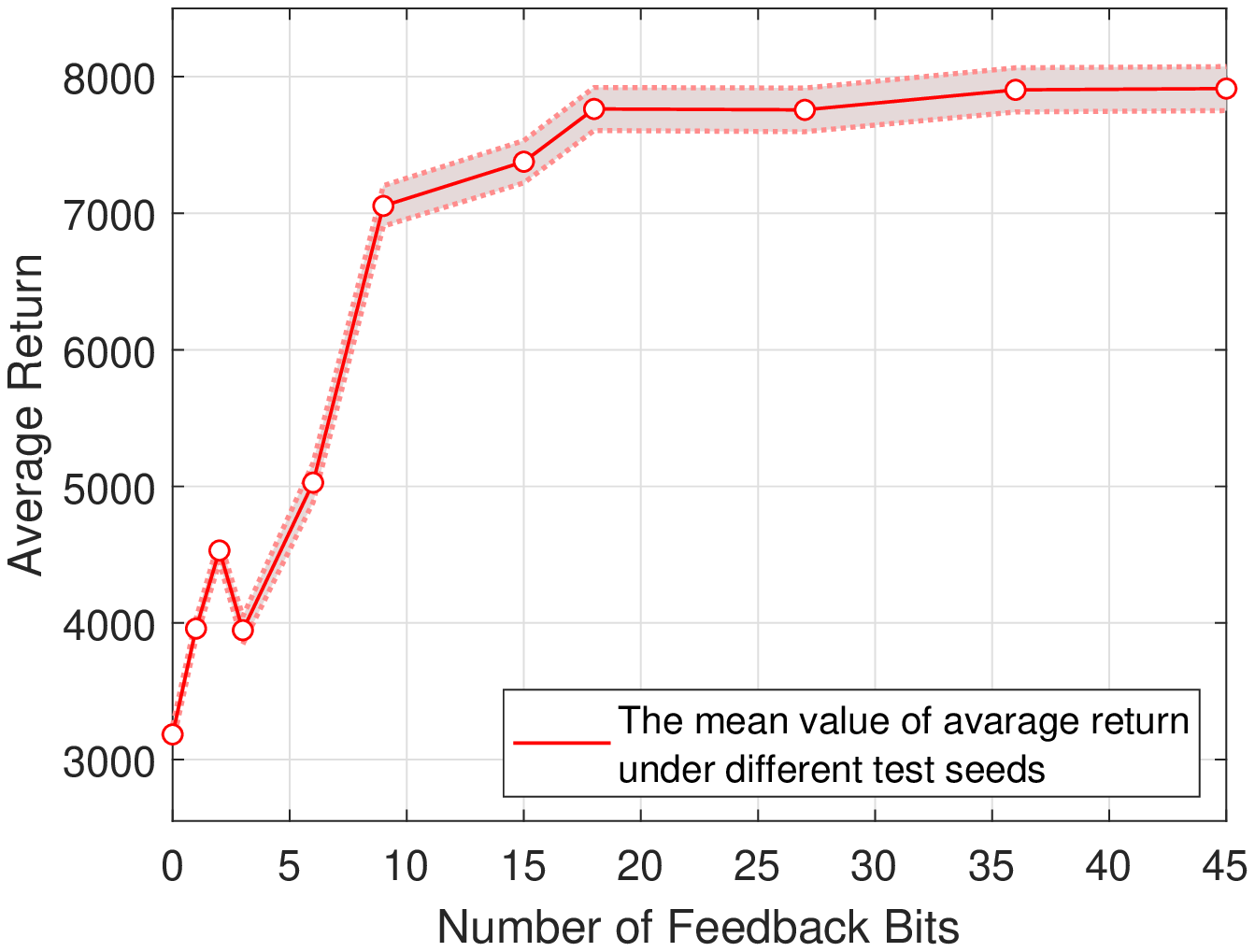}
\caption{Average return vs testing seeds}
\label{Fig7a}
\end{minipage}
\begin{minipage}[t]{0.45\textwidth}
\centering
\includegraphics[width=\textwidth]{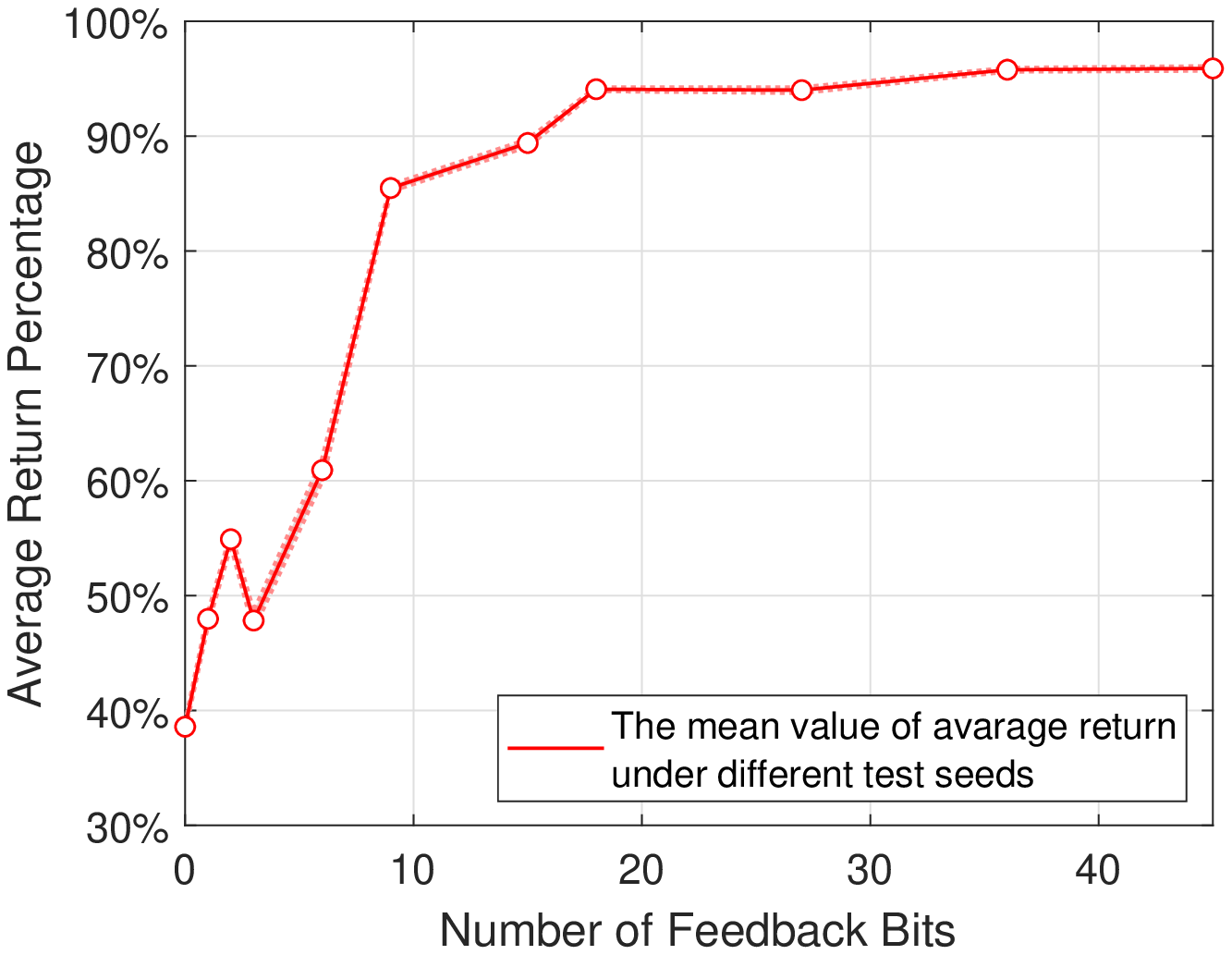}
\caption{ARP vs testing seeds}
\label{Fig7b}
\end{minipage}
\end{figure}

Fig. \ref{Fig6} demonstrates the change of the ARP with the increasing number of feedback bits under different mini-batch sizes $\mathcal{D}$. Here, we fix the real feedback number as $3$ and quantize each real feedback value into different numbers of feedback bits. In Fig. \ref{Fig6}, that the number of feedback bits equals $0$ refers to the situation where V2V links feed nothing back to the BS and just adopts the random action scheme.  The ARP first increases quickly with the number of feedback bits and then keeps nearly unchanged with the further increase of feedback bits after the number of feedback bits is over $18$. The ARP under different mini-batch sizes $\mathcal{D}$ has quite similar performance. Besides, the ARP can reach $95\%$ with $36$ feedback bits under $\mathcal{D}=512$. To tradeoff the number of feedback bits and the corresponding performance, we choose $36$ feedback bits under $\mathcal{D}=512$ in the subsequent evaluation.

The ARP fluctuates quickly when the number of feedback bits is smaller in Fig. \ref{Fig6}. This is mainly due to the sensitivity to different batch sizes or testing sequences under a small number of feedback bits. To further study this phenomenon, we depict the average return and the ARP under $10$ different testing seeds in Fig. \ref{Fig7a} and \ref{Fig7b}, respectively. Here, we choose $36$ feedback bits and $\mathcal{D}=512$. The achieved average return indeed varies under different testing seeds when the number of feedback bits is small in Fig. \ref{Fig7a}. When the number of feedback bits becomes larger, the mean values of average return under different testing seeds keeps increasing and then nearly unchanged with further increase of feedback bit. Besides, the variations of the average return under different testing seeds are constrained to a narrow range. Fig. \ref{Fig7b} depicts the variation of the ARP, where the average return under each testing seed is normalized by the average return of the optimal scheme under this seed. In Fig. \ref{Fig7b}, the ARP appears a very slight change under different testing seeds. Especially, the ARP has a wider range of variations when the number of feedback bits is quite small.

\begin{figure}[htbp]
\centering
\begin{minipage}[t]{0.45\textwidth}
\centering
\includegraphics[width=\textwidth]{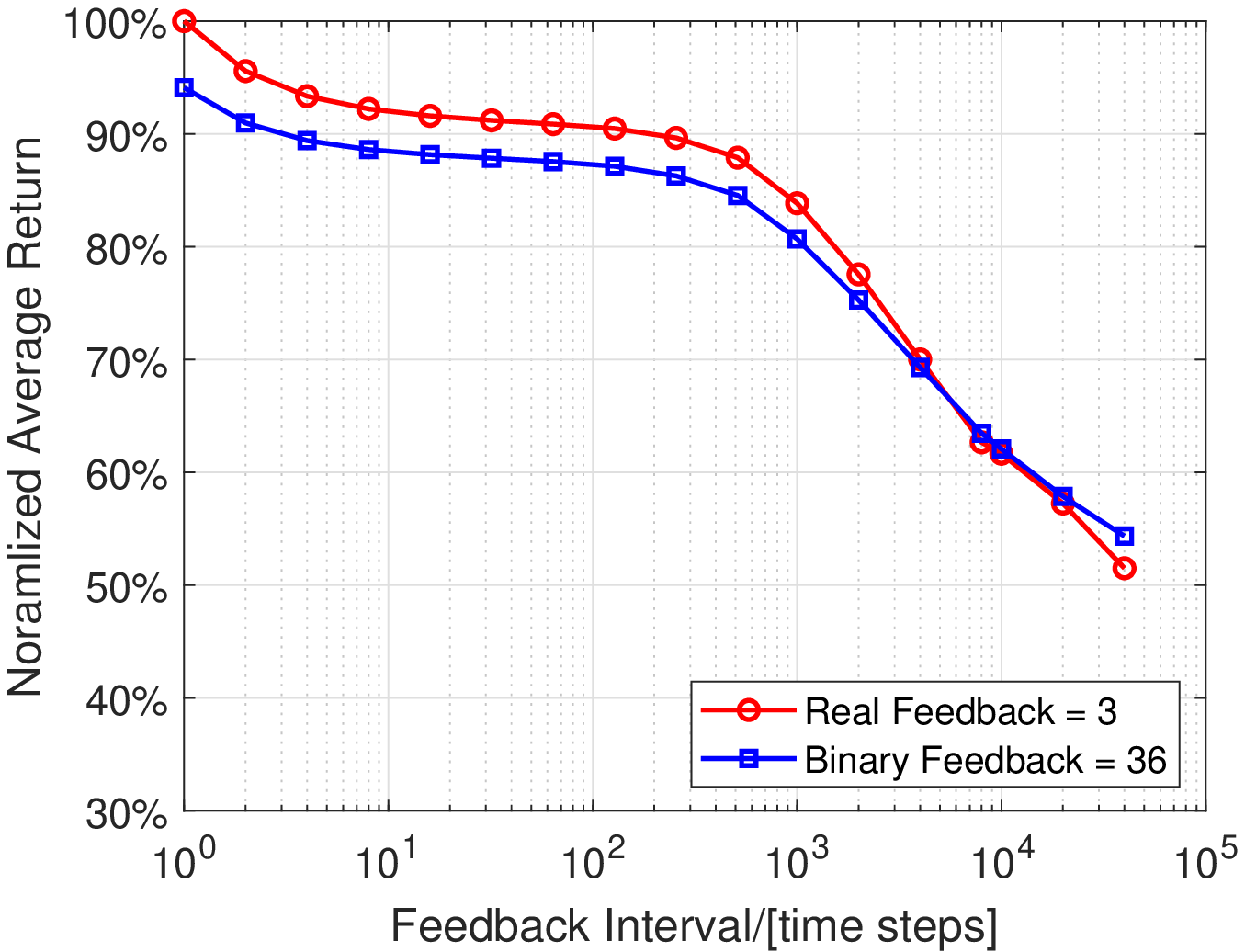}
\caption{Normalized Return vs feedback interval}
\label{Fig8}
\end{minipage}
\begin{minipage}[t]{0.45\textwidth}
\centering
\includegraphics[width=\textwidth]{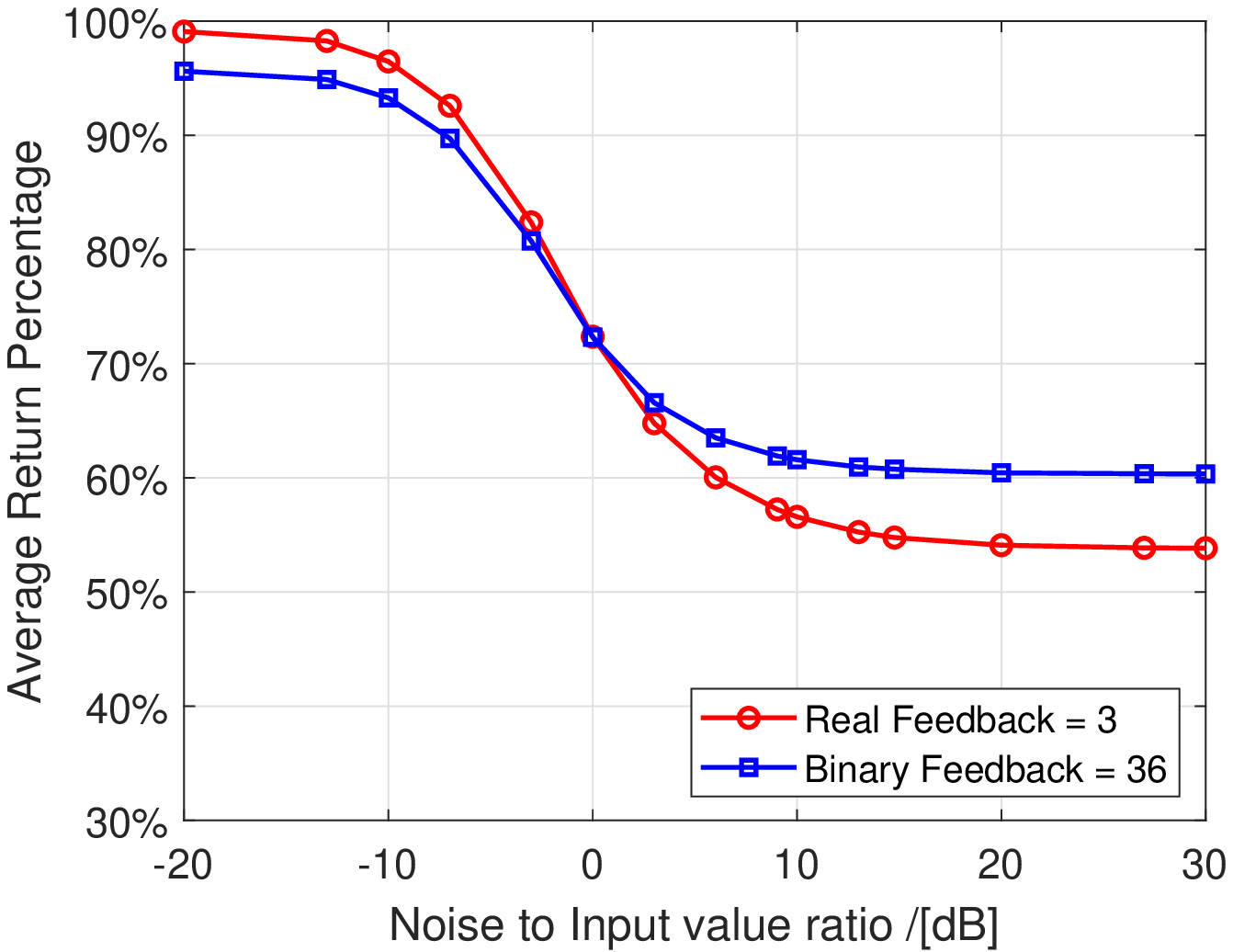}
\caption{ARP vs noisy input}
\label{Fig9}
\end{minipage}
\begin{minipage}[t]{0.45\textwidth}
\centering
\includegraphics[width=\textwidth]{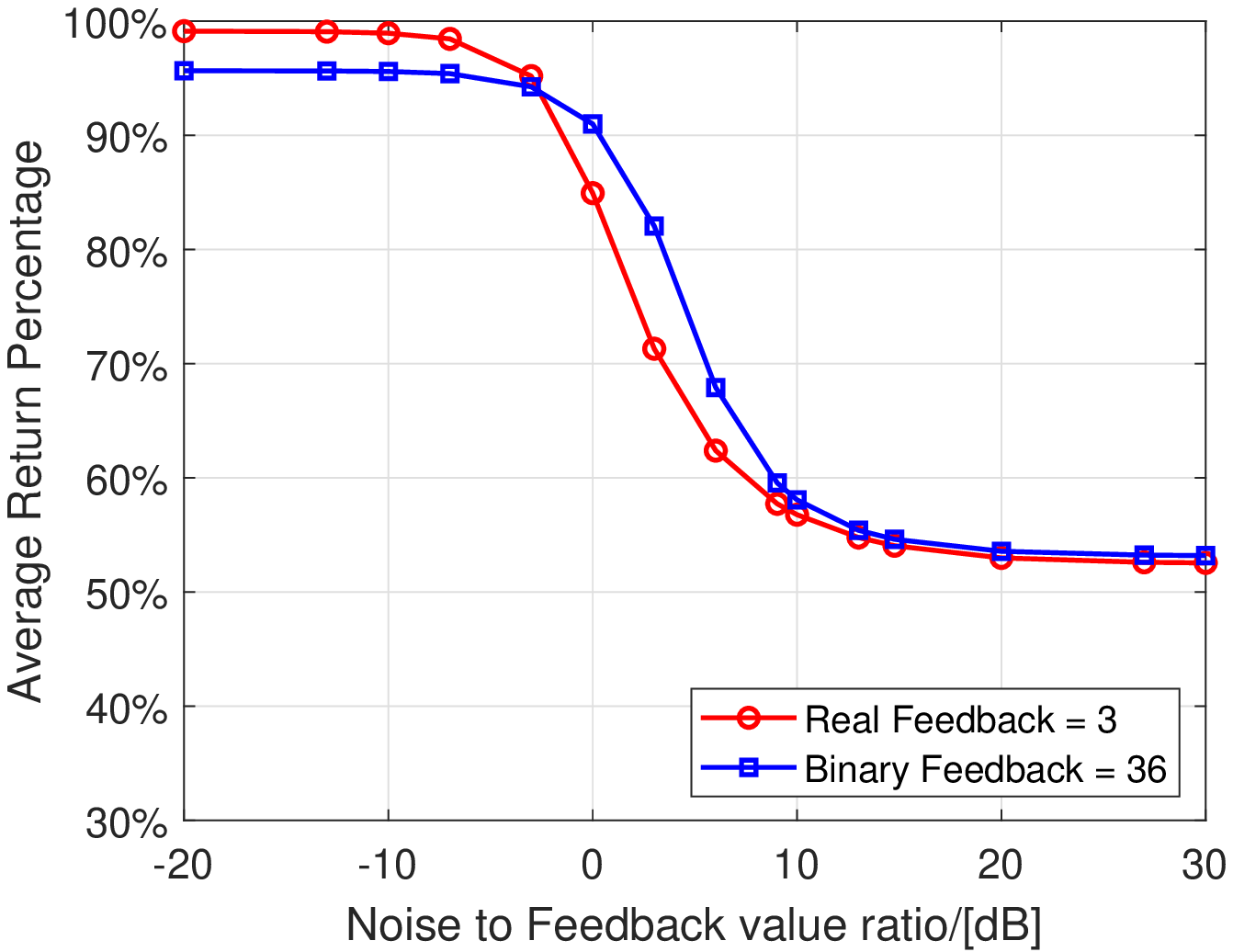}
\caption{ARP vs noisy feedback}
\label{Fig10}
\end{minipage}
\end{figure}

Fig. \ref{Fig8} shows the impacts of different feedback intervals on the performance of both real feedback and binary feedback, where the feedback interval is measured in the number of testing steps. From Fig. \ref{Fig8}, the normalized average return (where the average return is normalized by the average return under the scheme with $3$ real feedback since we set $T=50,0000$ and it is very high computational demanding to find the return under the optimal scheme) under both feedback schemes decreases quite slowly with the increasing feedback interval at the beginning, and then drops quickly with the very large feedback interval. Fig. \ref{Fig8} shows that the proposed scheme is immune to the feedback interval variations.

Fig. \ref{Fig9} illustrates the impacts of noisy input on the performance of both real feedback and binary feedback. Here, the x-axis means the ratio of the Gaussian white noise with respect to the specific value of each observation information (such as channel gain value) for V2V links. In addition, this ratio is expressed in dB to show a quite large range of input noise variation. In Fig. \ref{Fig9}, the ARP under both real feedback and binary feedback decreases very slowly at the beginning and then drops very quickly, and finally keeps nearly unchanged with the very large input noise, which shows the robustness of the proposed scheme. In addition, the proposed model can also gain $50\%$ and $60\%$ of the optimal performance under real feedback and binary feedback, respectively even at very large input noise, which is still better than the random action scheme. Besides, the binary feedback achieve better performance under large input noise than real feedback because the number of binary feedback is quite larger than that of the real feedback.

Fig. \ref{Fig10} displays the impacts of noisy feedback on the performance of both real feedback and binary feedback. Here, noisy feedback refers to the situation where noise inevitably occurs when each V2V link sends its learned feedback to the BS. Similarly, the x-axis means the ratio in dB of the Gaussian white noise with respect to the specific value of each feedback . In Fig. \ref{Fig10}, the ARP of both feedback schemes keeps nearly unchanged with the increasing feedback noise, which demonstrates the robustness of the proposed scheme, and then decreases more quickly under the real feedback compared with that under the binary feedback with the further increasing feedback noise. This is because there is only $3$ real feedback values under the real feedback scheme while $36$ feedback bits under the binary feedback scheme. Thus, the feedback noise has more impact on the real feedback when the feedback noise keeps increasing. Finally, the ARP of both feedback schemes becomes nearly constant with very large feedback noise. Similarly, the binary feedback is more robust to the feedback noise compared with the real feedback. However, even when the feedback noise becomes very large, such as $30$ dB, the ARP under both feedback schemes can still be bigger than $50\%$, which also shows the robustness of the proposed scheme to the feedback noise. That is, the proposed scheme can still achieve better performance than the random action scheme even when the feedback noise is very large. From another perspective, the proposed scheme can learn the intrinsic structure of the resource allocation in the V2X scenario.

\section{Conclusion}
In this paper, we proposed a novel architecture to allow the distributed V2V links to share spectrum efficiently through a scheme in which each D2D pair learns to feed channel and interference related information distributively to the BS that makes decisions for channel selection of D2D pairs. Each V2V link can learn what to feed back while the decision is made at the BS, which can achieve the near-optimal performance. To further reduce the feedback overhead and facilitate the transmission in the practical systems, we devise an approach to quantize the continuous feedback. From our simulation results, the quantization of the feedback performs reasonable well with an acceptable number of bits and our proposed scheme is quite immune to the variation of feedback interval, input noise, and feedback noise.

\bibliography{VtoVRef}



%
%
%
%

\end{document}